\documentclass[cameraready]{Interspeech}

\usepackage{amsmath,graphicx,hyperref}
\usepackage{enumitem}
\usepackage{booktabs}   
\usepackage{multirow}   
\usepackage{array}      
\usepackage{makecell} 
\usepackage[most]{tcolorbox}
\usepackage{colortbl}
\definecolor{tablebg}{RGB}{245,245,245} 

\usepackage{xcolor}
\usepackage{soul} 

\title{Adaptive Turn-Taking for Real-time Multi-Party Voice Agents}

\author[affiliation={1},]{Soumyajit}{Mitra}
\author[affiliation={1},]{Prabhat}{Pandey}
\author[affiliation={1},]{Abhinav}{Jain}
\author[affiliation={2},]{Shanmukha}{Sahith}
\author[affiliation={1},orcid=0000-0002-3332-1500]{K V Vijay}{Girish}

\address{
    $^1$ Amazon AGI \\
    $^2$ IIT Kharagpur, India 
}

\email{ssomit@amazon.com, panprabh@amazon.com, jabhinv@amazon.com, kvvijayg@amazon.com}


\usepackage{comment}

\begin{document}
\ninept

\maketitle

\begin{abstract}
Turn-taking in multi-party spoken conversations remains a fundamental challenge for voice-based agents, particularly under dynamic floor competition and varying user expectations. We propose \emph{ModeratorLM}, a role-playing voice agent that conditions turn-taking behavior on an explicitly assigned role in multi-party settings. The system is built on a speech large language model operating in chunk-wise streaming manner. We further introduce a reasoning-augmented variant that incorporates chain-of-thought reasoning over conversational context and the assigned role. We construct \emph{RolePlayConv}, a large-scale synthetic dataset of spoken multi-party conversations with diverse assistant roles. Experiments on real-world meeting data and \emph{RolePlayConv} show improved turn-taking precision by over 40\% and recall by more than 70\%, while substantially reducing false-positive interruptions compared to non-role-conditioned baselines.
\end{abstract}

\section{Introduction}
\label{sec:intro}

Recent advances in large language models (LLMs) have driven rapid progress in the development of voice-based conversational agents \cite{wang2024freeze,xie2024mini,chen2025minmo,ding2025kimi}. Modern spoken dialogue systems typically combine low-latency streaming speech processing modules with a core conversational component responsible for dialogue management and response generation. The emergence of full-duplex agents, capable of listening and speaking simultaneously, has further advanced voice-based conversational systems by enabling more natural interaction patterns such as backchannel responses and user barge-ins \cite{defossez2024moshi,ma2025language,yu2024salmonn}.
In dyadic (two-party) conversations, turn-taking is commonly governed by acoustic and semantic cues such as pause duration, silence detection, and sentence completion markers. Recent benchmarks have formalized the evaluation of turn-taking behavior in dyadic settings \cite{arora2025talking,lin2025full}. However, these assumptions do not generalize well to multi-party conversations, where interactions involve overlapping speech, dynamic floor competition, and negotiated turn allocation among participants \cite{sapkota2025multi}. In such settings, a conversational agent must continuously decide not only \emph{when} to speak, but also \emph{whether} to intervene at all.
At the same time, user expectations for conversational agents vary across contexts, ranging from passive listeners to active facilitators \cite{zhou2020design}. This has motivated research on role-playing language agents (RLPA), where assistants adopt explicit personas or roles \cite{chen2024persona,wang2024rolellm}. Prior work has primarily focused on text-based dialogue settings, modeling linguistic style \cite{zhou2024characterglm}, character traits \cite{ran2024capturing}, and decision-making patterns \cite{xu2024character}. However, the effect of an assigned role on real-time turn-taking behavior in voice-based, multi-party interaction has been largely overlooked.

In this work, we introduce \emph{ModeratorLM}, a role-playing voice agent for multi-party conversations that conditions both turn-taking decisions and response generation on an explicitly assigned role. Our system is built on a speech LLM that processes multi-speaker audio streams and corresponding text hypotheses in chunk-wise streaming manner. To our knowledge, this is the first role-conditioned voice agent designed for multi-party conversational settings. To support training, we construct a large-scale synthetic corpus of multi-party conversations spanning diverse assistant roles using off-the-shelf LLMs and a text-to-speech (TTS) model. Experimental results show that \emph{ModeratorLM} achieves substantially better alignment with the assigned roles than non-role-conditioned baselines, improving both role-consistent turn-taking and response generation. We further introduce \emph{ModeratorLM-Think}, which incorporates chain-of-thought reasoning \cite{wei2022chain} over conversational context and the assigned role, yielding additional improvements in in role-aligned turn-taking behavior.

\begin{figure*}[ht]  
    \centering
    \includegraphics[width=0.9\textwidth]{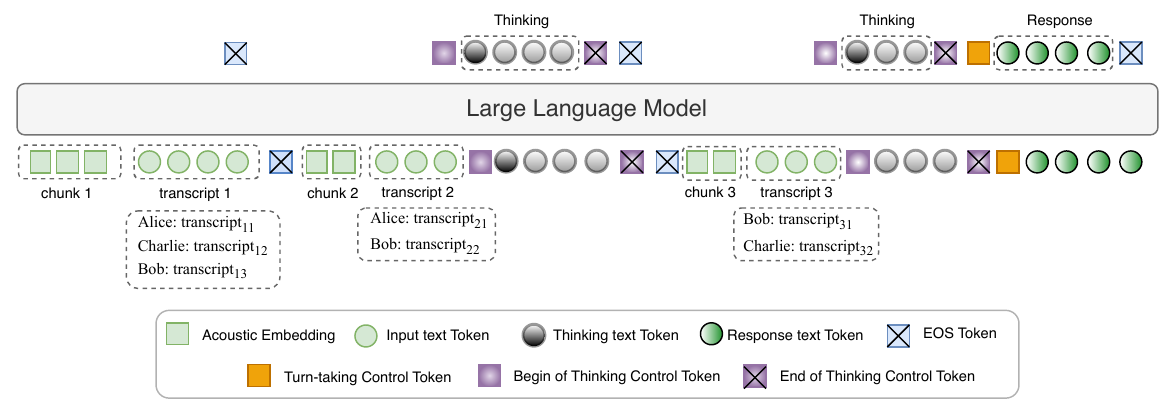}  
    \vspace{-2mm}
    \caption{Example input--output sequence of the LLM for ``ModeratorLM-Think'' model. No reasoning trace is produced in Chunk~1. A reasoning trace appears in Chunk~2 without turn-taking, while in Chunk~3 the assistant takes the floor.}
    \label{fig:speech_llm}
\vspace{-3mm}
\end{figure*}

\vspace{-1mm}
\section{Methodology}

\subsection{ModeratorLM: System Architecture}
\label{sec:system}

\textit{ModeratorLM} consists of a speech encoder and a backbone LLM. The speech encoder processes each incoming audio chunk independently and produces chunk-level embeddings. These embeddings are projected into the LLM embedding space via a trainable linear projection layer, following prior work \cite{pandey2025sift,jia2025efficient}. The multi-channel audio inputs are downmixed into a single-channel signal before encoding.
To support real-time interaction, chunk-level speech embeddings are appended sequentially to the LLM context in a streaming fashion, similar to \cite{ding2025kimi,zhang2024omniflatten,xu2025qwen2}. However, unlike approaches that rely on a separate voice activity detection module to determine turn boundaries \cite{ding2025kimi}, our system delegates turn-taking decisions entirely to the speech LLM itself. To improve robustness to variability in chunk duration, we adopt dynamically sized training chunks. For each chunk, the corresponding text transcription with speaker annotations is provided alongside the speech embeddings as input to the LLM.

For each input audio chunk, the LLM is trained to produce one of two output types:
(i) \textit{Turn-taking + Response}, where the model emits a control token indicating that the assistant should take the floor, followed by the assistant’s textual response; or
(ii) \textit{No-turn}, represented by an empty sequence, indicating that no turn-taking action is taken for the current chunk.
In addition, we train a reasoning variant, referred to as \textit{ModeratorLM-Think}, in which the LLM performs chain-of-thought reasoning at potential turn-taking points before emitting its decision. Figure~\ref{fig:speech_llm} illustrates an example input-output sequence for the LLM of \textit{ModeratorLM-Think} model.

Since this work focuses on modeling turn-taking behavior, the assistant’s responses are generated in text form rather than speech codes. A streaming TTS module (e.g., \cite{wang2024freeze}) can be integrated to produce real-time spoken responses, or native streaming speech token generation can be incorporated directly into the LLM following approaches \cite{xie2024mini,chen2025minmo}.

\vspace{-1mm}
\subsection{RolePlayConv Dataset Construction}
\label{sec:synth_data}

Most existing conversational datasets with explicitly defined assistant roles are text-based \cite{xu2024character,wang2025coser}. In the spoken domain, MELD \cite{poria2019meld} provides multi-party dialogue recordings but lacks assistant-role conditioning. More recently, \cite{fu2025pachat} introduced a persona-based multi-party conversational dataset; however, it focuses on personalized voice agents rather than role-conditioned assistants. Consequently, datasets that jointly support spoken multi-party conversations with role-conditioned assistants remain limited. To address this gap, we construct \textit{RolePlayConv}, a large-scale synthetic multi-party conversational dataset for role-playing voice agents.

The dataset is generated through a multi-stage pipeline. We first curate 125 detailed assistant roles (e.g., ``a 42-year-old Indian CEO with a confident and assertive communication style who enjoys strategic planning and mentoring''), specifying attributes such as conversational style and tone. For each conversation, an assistant role is randomly sampled and used to condition dialogue generation.
We then employ Amazon Nova Pro \cite{intelligence2024amazon} to generate coherent multi-party conversations spanning three to six speakers. Topic and subtopic sampling is constrained to align with the selected role to ensure consistency between role attributes and conversational content. To better reflect spoken interaction dynamics, individual dialogue turns are limited to fewer than 15 words, encouraging concise and conversational exchanges.
After generating the base dialogues, we augment all assistant turns and a subset of non-assistant turns with LLM-generated reasoning traces. These traces describe intermediate decision-making processes, including stance selection, response planning, and turn-taking considerations. They serve as supervision signals for training the reasoning-augmented variant, \textit{ModeratorLM-Think}. Table~\ref{tab:example} presents an illustrative example.

Finally, the textual conversations are synthesized into speech. Each speaker is assigned attributes such as gender, age group, and accent, which are mapped to a pool of reference voices. To avoid speaker leakage, distinct speaker pools are used for training and evaluation splits. Speech is synthesized turn-by-turn using the Zonos-v0.1 TTS model \cite{Zonos2025}. The resulting audio segments are concatenated with silence intervals sampled from a distribution estimated from real conversational data to approximate natural turn-level timing and pacing.
In total, \emph{RolePlayConv} contains approximately 75K training conversations, each averaging roughly two minutes in duration.

\begin{table}[t]
\centering
\footnotesize 
\rowcolors{1}{tablebg}{tablebg} 
\begin{tabular}{>{\raggedright\arraybackslash}p{0.95\linewidth}}
\toprule
\textbf{Role:} A 21-year-old Indian literature student with a soft, thoughtful voice who enjoys reading novels, sketching in quiet cafes, and listening more than speaking.\\
\midrule
\textbf{Conversation:} \\
\texttt{Hiya:} Hey everyone, I've been thinking a lot about patience and endurance lately. \\
\texttt{Lily:} Yeah, me too. It's tough, especially when you're excited to share something. \\
\texttt{[Thoughts]:} Lily is responding to Hiya's point about patience. I'll continue listening to build a clearer understanding of the group's perspective before speaking. \\
\texttt{Elise:} It's all about respecting others' space to speak, isn't it? Allowing one to finish can make conversations so much smoother. \\
\texttt{Pam:} I've noticed that when I let others finish, I understand them better too. \\
\texttt{[Thoughts]:} Pam just highlighted better comprehension through listening. The group's agreement feels settled, so I will add a concise thought linking patience to empathy without disrupting flow. \\
\texttt{Assistant:} Patience in listening fosters deeper understanding and connection.\\
\bottomrule
\end{tabular}
\vspace{0.5mm}
\caption{Sample conversation from RolePlayConv dataset involving four speakers and the assistant.}
\label{tab:example}
\vspace{-9mm}
\end{table}

\begin{table*}[ht]
\centering
\begin{tabular}{l | *{6}{c} | *{6}{c}}
\toprule
\multirow{2}{*}{\textbf{Model}} &
\multicolumn{6}{c|}{\textbf{NOTSOFAR-1}} &
\multicolumn{6}{c}{\textbf{RolePlayConv}} \\
\cmidrule(lr){2-7} \cmidrule(lr){8-13}
 & \textbf{@P} & \textbf{@R} & \textbf{@F1} & \textbf{@A} & \textbf{@FP} & \textbf{@RM}
 & \textbf{@P} & \textbf{@R} & \textbf{@F1} & \textbf{@A} & \textbf{@FP} & \textbf{@RM} \\
\midrule
Moshi &
0.14 & 0.10 & 0.11 &
0.21 & 0.66 & -- &
0.15 & 0.34 & 0.21 &
0.50 & 0.47 & -- \\

MP-Baseline &
0.58 & 0.33 & 0.38 &
0.69 & 0.05 & -- &
0.40 & 0.48 & 0.42 & 0.67 & 0.14 & -- \\
\midrule
ModeratorLM &
0.77 & 0.51 & 0.57 &
0.77 & 0.01 & 0.08 &
0.71 & 0.57 & 0.61 & 0.76 & 0.05 & 0.14 \\
ModeratorLM-Think &
0.81 & 0.74 & 0.76 & 0.86 & 0.01 & 0.02 &
0.79 & 0.82 & 0.79 & 0.91 & 0.03 & 0.03 \\
\bottomrule
\end{tabular}
\vspace{0.5mm}
\caption{Performance comparison across different models on NOTSOFAR-1 and RolePlayConv test sets.
@P, @R, @F1, @A, @FP, and @RM denote precision, recall, F1-score, macro-accuracy, false-positive rate, and reactive miss rate, respectively. Reactive miss rate (@RM) is not applicable for Moshi and MP-Baseline, as these models do not support role configuration.}
\label{tab:main}
\vspace{-8mm}
\end{table*}

\vspace{-1mm}
\section{Experimental Setup}
\label{sec:experiments}

\subsection{Training Setup}
\label{sec:train_setup}

We use \textit{Qwen3-4B-Instruct-2507} \cite{yang2025qwen3} as the backbone LLM for \textit{ModeratorLM}, and \textit{Qwen3-4B-Thinking-2507} for \textit{ModeratorLM-Think}. For speech representation, we employ an in-house speech encoder trained with variable lookahead similar to \cite{swietojanski2023variable, male2025durepdualmodespeechrepresentation}, enabling block-wise attention during the inference on variable-sized chunks. To construct streaming inputs, we obtain word--time alignments using the Montreal Forced Aligner \cite{mcauliffe2017montreal}. These alignments are used to derive chunk boundaries for training.
Because chunking is typically controlled by an external module independent of the conversational model, we adopt a dynamic chunking policy during training. Specifically, chunk lengths are randomly sampled between 0.5\,s and 3\,s, improving the robustness to diverse chunking behaviors. We additionally ensure that a fraction of chunks terminate at speaker boundaries, as these correspond to natural turn-taking points.

Our training pipeline consists of three stages:

\noindent\textbf{Speech–LLM Alignment.}
We first align the speech embeddings with the LLM input space by training on automatic speech recognition (ASR) task using approximately 90K hours of public speech data (VoxPopuli \cite{wang2021voxpopuli}, MLS \cite{pratap2020mls}, Common Voice \cite{ardila2019commonvoice}, People's Speech \cite{galvez2021peoplespeech}). During this stage, only the projection layer is updated, while all other model parameters remain frozen.

\noindent\textbf{Conversation Pretraining.}
We then train on a mixture of publicly available multi-party conversation datasets (AMI \cite{carletta2005ami}, Fisher \cite{cieri2004fisher}). Since public corpora do not have a notion of ``assistant'' speaker, we simulate one by rotating the assistant role among all speakers except the conversation initiator. For a conversation with $N$ participants, this yields $N-1$ training instances.

\noindent\textbf{Role-Conditioning Training.}
In the final training stage, we fine-tune exclusively on the \emph{RolePlayConv} (Section~\ref{sec:synth_data}) dataset, where the assistant's role is specified via the system prompt.

For the latter two stages, we fine-tune the LLM parameters using low-rank adaptation \cite{hu2021lora}, while keeping the speech encoder frozen throughout all stages (13.4M trainable parameters).
We use the Adam optimizer with a two-phase learning rate schedule with the peak learning rate of $1 \times 10^{-5}$.

\vspace{-1mm}
\subsection{Evaluation Setup}
\label{sec:eval_setup}

\noindent\textbf{Evaluation Datasets.}
We evaluate our models on two multi-party spoken conversation datasets.
The first is NOTSOFAR-1 \cite{vinnikov2024notsofar} (\emph{NSF-1}), which consists of real recordings of formal meetings with approximately four speakers per session and an average duration of six minutes.
As \emph{NSF-1} lacks role labels, we designate one speaker as the ``assistant'' using a hybrid procedure that aggregates LLM-based rankings of assistant-like behavior with independent human evaluations. A role description is then generated for the speaker selected as assistant.

The second evaluation set is constructed following the same data generation pipeline as \textit{RolePlayConv} (Section~\ref{sec:synth_data}). To ensure a strict separation from training conditions, we use a zero-shot set of roles unseen during training and a different LLM, \texttt{QwQ-32B} \cite{team2025qwq}, for the evaluation set construction.

\noindent\textbf{Inference Configuration.}
During evaluation, each incoming audio chunk is treated as a potential turn-taking point. We adopt a dynamic chunking scheme, where chunk durations are uniformly sampled between 0.5\,s and 3\,s, and segmentation always occurs at speaker boundaries. As dynamic chunking introduces non-determinism, we report metrics averaged over 10 evaluation runs. In Section~\ref{sec:ablation}, we show that fixed-size chunking is not suitable for turn-taking evaluation.
To obtain turn-taking predictions, we teacher-force the ground-truth conversation context together with the audio chunk under evaluation before invoking generation. Greedy decoding is used by default. When a thinking control token (\texttt{<think>}) is emitted, we switch to sampling mode to generate the reasoning tokens. Similarly, upon emitting a turn-taking control token, sampling is used to generate the assistant’s response. Sampling is performed with \emph{Temperature=0.7}, \emph{TopP=0.8}, and \emph{TopK=20}.

\noindent\textbf{Evaluation Metrics.}
We evaluate predicted turn-taking decisions against ground-truth annotations at the chunk level using standard binary classification metrics: Precision, Recall, F1-score, Macro-averaged Accuracy, and False Positive Rate. In addition, we report the \textit{Reactive Miss Rate}, which measures the fraction of missed turn-taking opportunities where a speaker directly addresses the assistant.
To account for the inherent subjectivity of turn-taking and role adherence, we also conduct LLM-as-a-Judge evaluation \cite{zheng2023judging} using \emph{Claude-Sonnet-3.5} \cite{anthropic2024claude35sonnet}. The judge model is provided with the assistant’s role description and the conversation history, and is prompted to assess (i) the appropriateness of the turn-taking decision with respect to the assigned role on a 0--1 scale, and (ii) role fidelity of the response content on a 1--10 scale. A small-scale human evaluation conducted on 100 instances shows strong agreement between human judgments and LLM-based scores, with a Spearman rank correlation of $\rho = 0.87$.

\noindent\textbf{Baselines.}
We compare against two non-role-conditioned voice agent baselines that do not support role conditioning. The first is Moshi \cite{defossez2024moshi}, designed for dyadic conversations (we use \emph{Moshika}, the female-voice variant). Because Moshi operates at a frame rate of 12.5\,Hz and produces continuous streaming outputs, we evaluate its turn-taking predictions using a tolerance window of $\pm$0.5\,s around the ground-truth turn-taking points.
The second baseline is the model fine-tuned on the \emph{RolePlayConv} dataset without role-conditioning after the \emph{Conversation Pretraining} stage (Section~\ref{sec:train_setup}), which we refer to as \textit{MP-Baseline}. As this model is trained on the same dataset as \emph{ModeratorLM} models but without role-conditioning signals, it allows us to isolate the effect of role-conditioning on turn-taking behavior.

\vspace{-2mm}
\section{Results}
\label{sec:results}

\subsection{Main Results}
\label{sec:main_results}



Table~\ref{tab:main} compares \emph{ModeratorLM} variants against non-role-conditioned baselines on \emph{NSF-1} and \emph{RolePlayConv} datasets. \emph{Moshi}, trained on dyadic conversations, fails to generalize to multi-speaker settings, exhibiting very low recall and high false positive rates. The \emph{MP-Baseline}, trained on multi-party conversations but without role conditioning, reduces interruptions but lacks the precision required for role-conditioned behavior.
Both non-role-conditioned models exhibit substantially lower recall on \emph{NSF-1}, which contains natural human-to-human conversations with frequent backchannels, interruptions, and overlapping speech. In contrast to the more structured user--assistant interactions in \emph{RolePlayConv}, no participant consistently follows assistant-style turn-taking patterns in \emph{NSF-1}.

The \emph{ModeratorLM} model variants perform substantially better across both datasets. \emph{ModeratorLM} achieves high precision with very low false positives but remains conservative, missing some valid response opportunities. Incorporating chain-of-thought reasoning in \emph{ModeratorLM-Think} improves this trade-off, yielding higher recall while maintaining low false positive rates, suggesting that explicit reasoning helps the model better interpret conversational context and role-specific obligations. Moreover, explicit reasoning reduces the miss rate at reactive turn-taking points.


Table~\ref{tab:subjective} presents the LLM-as-a-Judge evaluation of role fidelity, where subjective trends closely mirror objective metrics. \emph{ModeratorLM-Think} achieves the highest scores, indicating that the explicit reasoning stage serves as a foundational alignment layer. By first analyzing conversational context and role-specific obligations, the model ensures the decision of \emph{when to speak} is cognitively grounded. This structured dependency yields turns that are both accurately timed and stylistically consistent.
Finally, Table~\ref{tab:persona-context-thoughts} provides qualitative examples demonstrating how the model adapts its reasoning under different roles given the same conversational context, resulting in distinct turn-taking decisions.

\begin{table}
\centering
\begin{tabular}{l c c}
\toprule
\textbf{Model} & \textbf{Turn-Taking} & \textbf{Response} \\
\midrule
MP-Baseline  & 0.58 & 4.6 \\
\midrule
ModeratorLM       & 0.68 & 6.9 \\
ModeratorLM-Think & 0.72 & 7.4 \\
\bottomrule
\end{tabular}
\vspace{0.5mm}
\caption{LLM-as-a-judge evaluation of role fidelity for turn-taking ([0,1] scale) and responses ([0,10] scale) on RolePlayConv evaluation set.}
\label{tab:subjective}
\vspace{-6mm}
\end{table}

\begin{table}
\centering
\footnotesize
\rowcolors{1}{tablebg}{tablebg} 
\begin{tabular}{>{\raggedright\arraybackslash}p{0.95\linewidth}}
\toprule
\textbf{Roles:}
\vspace{0.5mm}
\begin{itemize}[nosep, leftmargin=*]
\item \textbf{Role A:} A 46-year-old American news anchor with a crisp, authoritative delivery who enjoys live reporting, moderating panels, and investigative journalism.
\item \textbf{Role B:} A 46-year-old American news anchor with a composed, non-intrusive demeanor who enjoys live reporting, moderating panels, and investigative journalism.
\vspace{-3.5mm}
\end{itemize}\\
\midrule
\textbf{Conversation Context:}\\
\texttt{Nisha:} Hey everyone, let's talk about leveraging influence and leadership.\\
\texttt{Lance:} Totally agree Nisha, it's crucial for driving change.\\
\midrule
\textbf{Thoughts:} 
\begin{itemize}[nosep, leftmargin=*]
\item \textbf{Role A:} Nisha initiated the topic, and Lance reinforced its importance. As a moderator in panels before, this discussion felt like an opening to set clear parameters for actionable insights without derailing their momentum into abstract theorizing. The first step is structure clarity through concise statements.  
\item \textbf{Role B:} Nisha initiated the topic, and Lance is affirming it. The conversation needs to build momentum before interjecting. Sometimes, silence allows ideas space without pressure from a passive anchor who values structured dialogue over interruption.  
\vspace{-3.5mm}
\end{itemize}\\
\bottomrule
\end{tabular}
\vspace{0.5mm}
\caption{Reasoning traces generated by \textit{ModeratorLM-Think} under two different roles for the same conversation context.}
\label{tab:persona-context-thoughts}
\vspace{-8mm}
\end{table}

\begin{table}
\centering
\setlength{\tabcolsep}{4pt} 
\begin{tabular}{l | *{3}{c}| *{3}{c}}
\toprule
\multirow{2}{*}{\textbf{Setup}} &
\multicolumn{3}{c|}{\textbf{ModeratorLM}} &
\multicolumn{3}{c}{\textbf{ModeratorLM-Th.}} \\
\cmidrule(lr){2-4} \cmidrule(lr){5-7}
 & @P & @R & @A & @P & @R & @A \\
\midrule
Default & 0.71 & 0.57 & 0.76 & 0.79 & 0.82 & 0.91 \\
\midrule
No Transcription & 0.42 & 0.14 & 0.57 & 0.39 & 0.42 & 0.57 \\
ASR Hypotheses & 0.68 & 0.56 & 0.76 & 0.75 & 0.80 & 0.90 \\
GT Thoughts & -- & -- & -- & 0.95 & 0.95 & 0.97 \\
\midrule
Fixed (2\,s) & 0.88 & 0.78 & 0.88 & 0.82 & 0.82 & 0.91 \\
Turn-Fixed & 0.84 & 0.60 & 0.80 & 0.75 & 0.81 & 0.91 \\
\bottomrule
\end{tabular}
\vspace{0.5mm}
\caption{Evaluation results on RolePlayConv under different setups. @P, @R, and @A denote precision, recall, and macro-averaged accuracy, respectively. GT denotes ground truth. The default setup uses a fully dynamic chunking policy.}
\label{tab:ablation}
\vspace{-8mm}
\end{table}

\vspace{-1mm}
\subsection{Ablation Studies}
\label{sec:ablation}

Table~\ref{tab:ablation} reports results under different evaluation configurations designed to analyze the impact of chunking strategies and transcription availability on turn-taking. We compare the default \emph{dynamic} chunking policy with two alternatives:
(i) \emph{Fixed}, where a constant chunk size is applied across all utterances; and
(ii) \emph{Turn-Fixed}, where the chunk size is constant within each speaker turn but may vary across turns, even within the same utterance.
We define an \emph{utterance} as a sequence of consecutive speaker turns occurring before the assistant responds, based on ground-truth annotations. For both alternative policies, segmentation is aligned with utterance boundaries, resulting in a variable-length final chunk for each utterance.

Under these settings, \textit{ModeratorLM} exhibits higher turn-taking precision with both alternative chunking strategies compared to dynamic chunking. In the \emph{Fixed} setting, recall also improves, as turn-taking chunks are typically shorter, while non–turn-taking chunks remain constant in length. This behavior suggests that \textit{ModeratorLM} relies substantially on chunk-size cues when making turn-taking decisions. In contrast, \textit{ModeratorLM-Think} is notably less sensitive to chunking strategy, instead relying on explicit reasoning traces. This robustness helps explain its near-perfect performance when ground-truth reasoning traces are provided.
However, the apparent gains from fixed-size chunking are partly an artifact of the evaluation setup. In realistic inference scenarios, ground-truth utterance boundaries are unavailable. A fixed chunking policy must process an entire chunk before passing it to the model, which can delay or obscure turn-taking cues that emerge mid-chunk, leading to missed or mistimed interventions.

We also study the effect of transcriptions of user turns. Specifically, we evaluate two settings:
(i) \emph{No Transcription}, where no textual input is provided; and
(ii) \emph{ASR Hypotheses}, where transcripts are obtained using the streaming Kyutai-STT-2.6B model \cite{zeghidour2025streaming} independently on each user audio channel (word error rate of 6.7\%).
Performance degrades substantially in the \emph{No Transcription} setting, indicating that the model relies heavily on textual information for turn-taking decisions. In contrast, using ASR hypotheses results in only minor degradation, demonstrating robustness to realistic transcription errors.

\vspace{-2mm}
\section{Conclusions}
\label{sec:conclusions}

In this work, we introduced a role-playing voice agent for multi-party conversations that modulates turn-taking behavior based on an assigned role. Experimental results show that role-conditioned fine-tuning yields turn-taking decisions better aligned with configured preferences, and that incorporating chain-of-thought reasoning further improves role fidelity. Our analysis also highlights the importance of dynamic chunking during training; without it, models tend to overfit to chunk length rather than conversational context. Future work may explore full-duplex multi-party voice agents and turn-taking under simultaneous listening and speaking.

\section{Acknowledgments}
\label{sec:acknowledgments}
We would like to thank Ajay Srinivasamurthy, Volker Leutnant, Adam Kaplan, Andreas Schwarz, Raghavendra Bilgi and Sri Garimella for their support and valuable feedback.

\section{Generative AI Use Disclosure}
The authors acknowledge the use of generative AI tools during the preparation of this paper strictly for the purposes of editing, polishing, and improving the readability of the manuscript. Generative AI was not used in the conceptualization, experimental design, or generation of the core scientific content. All co-authors take full responsibility and accountability for the entirety of the work and consent to its submission in accordance with ISCA policy.

\bibliographystyle{IEEEtran}
\bibliography{mybib}

\end{document}